# SIMULATIONS OF THE MICROWAVE SKY AND OF ITS "OBSERVATIONS"

F.R. BOUCHET and R. GISPERT

*Institut d'Astrophysique, Paris, and Institut d'Astrophysique Spatiale, Orsay, France*

and

N. AGHANIM, J.R. BOND, A. DE LUCA, E. HIVON and B. MAFFEI *

**Abstract.**
Here follows a preliminary report on the construction of fake millimeter and sub-millimeter skies, as observed by virtual instruments, e.g. the COBRA/SAMBA mission, using theoretical modeling and data extrapolations. Our goal is to create maps as realistic as possible of the relevant physical contributions which may contribute to the detected signals. This astrophysical modeling is followed by simulations of the measurement process itself by a given instrumental configuration. This will enable a precise determination of what can and cannot be achieved with a particular experimental configuration, and provide a feedback on how to improve the overall design. It is a key step on the way to define procedures for the separation of the different physical processes in the future observed maps. Note that this tool will also prove useful in preparing and analyzing current (*e.g.* balloon borne) Microwave Background experiments.

## 1. Introduction

The primary objective of the COBRA/SAMBA mission submitted to ESA and of the SAMBA mission submitted to CNES is to determine the spatial characteristics of the fluctuations of the Cosmological Microwave Background radiation (hereafter CMB), at all angular scales between five minutes of arc and a few degrees. In order to reach that goal with the required precision (detection of signals with an equivalent $\frac{\Delta T}{T} \sim 10^{-6}$, *i.e.* $\sim 3\mu K$), it is necessary to separate the numerous possible contributions to this radiation. We have embarked on the construction of simulated millimeter and submillimeter skies, using theoretical modeling and data extrapolations, which we then observe with virtual instruments, e.g. those on the COBRA/SAMBA payload. Our goal is to create maps as realistic as possible of the relevant physical phenomena which may contribute to the detected signals.

The usual approach of comparing the expected variances of the distribution of fluctuations of different types (*i.e.* their *rms* values) is not sufficient because these distribution functions are certainly non-Gaussian for some of the sources. For example, the submillimeter emission of resolved galaxies and the Sunyaev–Zeldovich effect due to clusters are certainly in that category

* NA, AdL, and BM are from the Institut d'Astrophysique Spatiale, Orsay, France; RJB is from CITA, Toronto, Canada; EH is from the Institut d'Astrophysique, Paris.

since there are localized objects. But this might also be true for other sources of fluctuations such as those arising from topological defects, remnants of an earlier phase in the early Universe. As another example, the emission of the dust itself at the maximum wavelength of the CBR, either from the Milky Way or from other galaxies, can probably be removed by spectral extrapolations from maps at shorter wavelengths. In that case, we wish to study the spatial properties of the residuals under various assumptions concerning, e.g. the respective distribution of a cold and hot dust component. Another reason is that the resulting maps may be used as test beds for ideas concerning the data analysis process. We hope to check then the feasibility of various signal-separation techniques. Furthermore, we plan to use this tool in an iterative fashion in order to optimize the instrumental characteristics of the planned experiment in order to best discriminate both spatially and spectrally the various components.

## 2. Sky Simulations

The sources of flux anisotropies which may contribute at the wavelengths of the instrument (say between $200\,\mu$m and $2\,$cm) can be decomposed as follows:

- Primary $\frac{\Delta T}{T}$ of the 2.726 K background. Those are imprinted predominantly during the last Thomson scatterings of photons by free electrons.
- Sunyaev-Zeldovich (SZ) effect from the hot gas in galaxy clusters. It may in turn be decomposed into 2 pieces:
  · $y$-distortions (thermal part), due to the scattering of "cold" CMB photons off the "hot" electron of the ionized intracluster gas
  · $\frac{\Delta T}{T}$ distortions (kinematical part), due to the Doppler shift from the clusters peculiar velocities accompanying Thomson scattering
- Submillimeter emission from the galaxies, which can also be decomposed into two pieces:
  · contribution from spatially resolved galaxies
  · background fluctuations due to the integrated emission of unresolved galaxies (including starburst galaxies and AGN's)
- Emission from our own Galaxy. Stellar sources have a weak contribution at the wavelength of interestm but there are three main other components which are likely to be relevant:
  · Interstellar dust (even at high galactic latitude: galactic cirrus)
  · Bremsstrahlung emission
  · Synchrotron radiation
- Other sources of fluctuations. Let us mention just a few: topological defects created during an early Universe Symmetry–breaking phase transition (*e.g.* Cosmic Strings, Textures, Global monopoles, etc...), or, in the Earth neighborhood, Zodiacal light or asteroïds trails.

At this preliminary stage, our modeling of the various physical phenomena has been done as follows:

- For the primary $\frac{\Delta T}{T}$ and the SZ effect from clusters (both $y$ & $\frac{\Delta T}{T}$) we have created realizations corresponding to specific theoretical models (*e.g.* standard CDM). This was also done for the possible secondary fluctuations from cosmic strings.
- For spatially resolved galaxies and the galactic dust, as a simple first step, we have split $100\,\mu$m ISSA-IRAS maps into two maps. The first one is a population of resolved extragalactic sources assumed to be at a uniform temperature $T = 30K$ (with a dust standard emissivity $\varepsilon_\nu \propto \nu^{-2}$). The second one is the dust galactic background which we assumed to be also at a uniform temperature, $T = 18K$.
- For the synchrotron radiation, we have simply extrapolated the 408 MHz maps from Haslam (1982) (using a $I_\nu \propto \nu^{-0.7}$ law).

All these processes can be stored as elementary maps (presently $12.5° \times 12.5°$, with $500^2$ pixels, each of size 1.5 arcminutes) together with conversion coefficients to energy flux density at a given frequency.

We are currently modeling the fluctuations from unresolved galaxies by Monte-Carlo simulations. We draw a population of galaxies by redshift interval. In each interval, we take into account the spatial correlations between objects and their luminosity functions, according to their (assumed) time–evolution. The spectra of the galaxies are determined from their luminosity and type. The results of this study are not included yet. We plan to use more precise models of dust emissivity as in Désert et al.(1990). And the $60\mu$m IRAS maps and DIRBE-COBE data at high latitude will furthermore permit us to actually deduce the dust temperature fluctuations. We also urgently need to include estimates of the Bremsstrahlung emission (for the COBRA wavelength), possibly using COBE maps or $H_\alpha$ maps extrapolated to the small angular scales of relevance to us.

## 3. Satellite Measurements Simulations

All of our virtual instruments have at this stage the following characteristics:
- nominal spectral bands with unity transmission across the entire band
- Gaussian lobes of FWHM equal to the diffraction limit
- a $1/f$ detector noise with a low frequency cutoff
- the orbital type is a simplified model assuming full sky coverage, but with integration time varying from 30s to 1000s per pixel.

The maps will also depend on the accuracy level retained for the digital data sent to the ground. One further issue will be the type, if any, of on–board compressing and pre-processing of the data. The structure of our tool is designed to deal easily with further modifications of the experimental configuration. Note that to avoid spurious effects on the boundaries, we keep only the inner $10° \times 10°$ of the maps, onced convolved with a lobe.

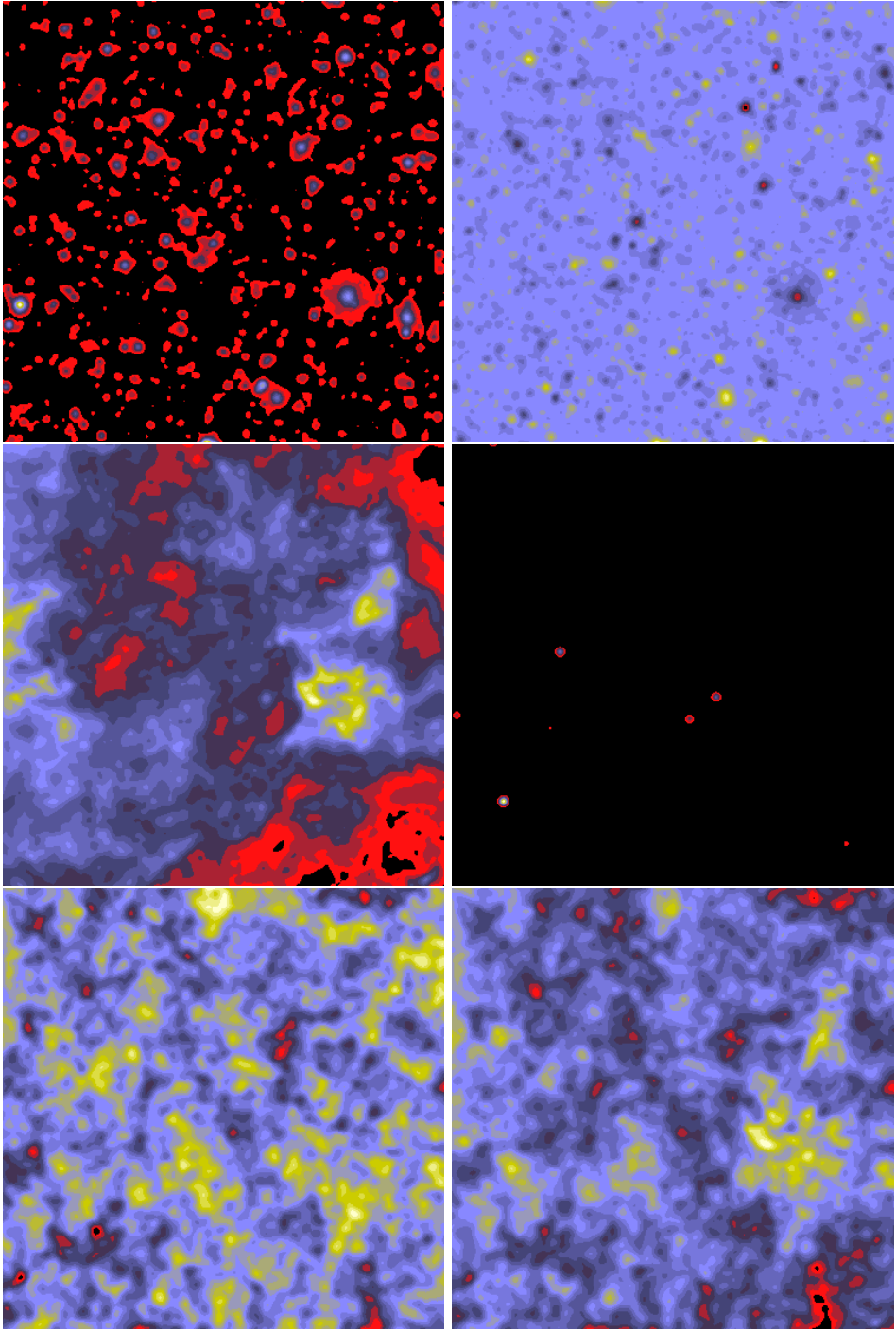

Fig. 1. $10 \times 10°$ maps in the 1.44 mm band for a gaussian lobe of 7.5' FWHM. They correspond, from left to right, and from top to bottom, to the thermal and kinematical SZ effect, to the dust and foreground galaxies, to CDM primary CMB fluctuations and the sum of all the above plus the (not shown) synchrotron contribution.

## 4. Sample Results and Perspectives

Figure 2 shows a possible choice of 5 contributing processes and their resulting sum plus an (not shown, but included) synchrotron component. The scales of the different maps are independant. They corresponds to a same band with uniform transmission of unity between 1.1 and 1.8 mm, and an observation with a gaussian beam of 7.5 minutes of arc. At this wavelength, the dominating fluctuations are those of the CMB. Still, it is clear that the pattern of the sum map is rather different from that of the CMB alone. See Table I for the contributions of the processes to the total variance in the other wavebands.

| Mean $\lambda$(mm) (& $\Delta\lambda/\lambda$) | 2.23 (0.4) | 1.44 (0.5) | 0.85 (0.7) | 0.46 (0.6) |
|---|---|---|---|---|
| FWHM (arc-min) | 10.5 | 7.5 | 4.5 | 3.0 |
| Primary $\Delta T/T$ | 93.174 | 53.748 | 0.839 | $\leq$0.001 |
| SZ ($y$) | 0.083 | $\leq$0.001 | 0.006 | $\leq$0.001 |
| SZ ($\Delta T/T$) | 0.002 | 0.002 | $\leq$0.001 | $\leq$0.001 |
| Foreground galaxies | $\leq$0.001 | $\leq$0.001 | 0.007 | 0.026 |
| Galactic dust | 6.737 | 46.249 | 99.148 | 99.973 |
| Synchrotron | 0.003 | $\leq$0.001 | $\leq$0.001 | $\leq$0.001 |

TABLE I

Contribution (in %) of each process to the total variance of the maps in each SAMBA channel. Note though that this is quite deceiving for non-gaussian contributions.

We already have available analyses like distributions of pixel intensities, or $C_\ell$ decompositions (*i.e.* the coefficients of a standard spherical harmonics decomposition), see Fig. 4. We are implementing measurements of the angular correlation function of the anisotropies, $C(\theta)$, and of higher order statistics. We shall soon start investigating effective signal–separation techniques.

Of course, our current modeling is far from perfect. For the high–$z$ sources of fluctuations, we have to rely on theoretical models. In some cases, years of effort have led to the development of reliable tools, such as for the primary $\Delta T/T$ fluctuations in Gaussian models. In other case, it is technically hard to make definite predictions, even though the model is perfectly well specified (e.g. for strings, or the SZ effect which depends on the gas behavior in very dense environments). We also have only weak constraints on the high–

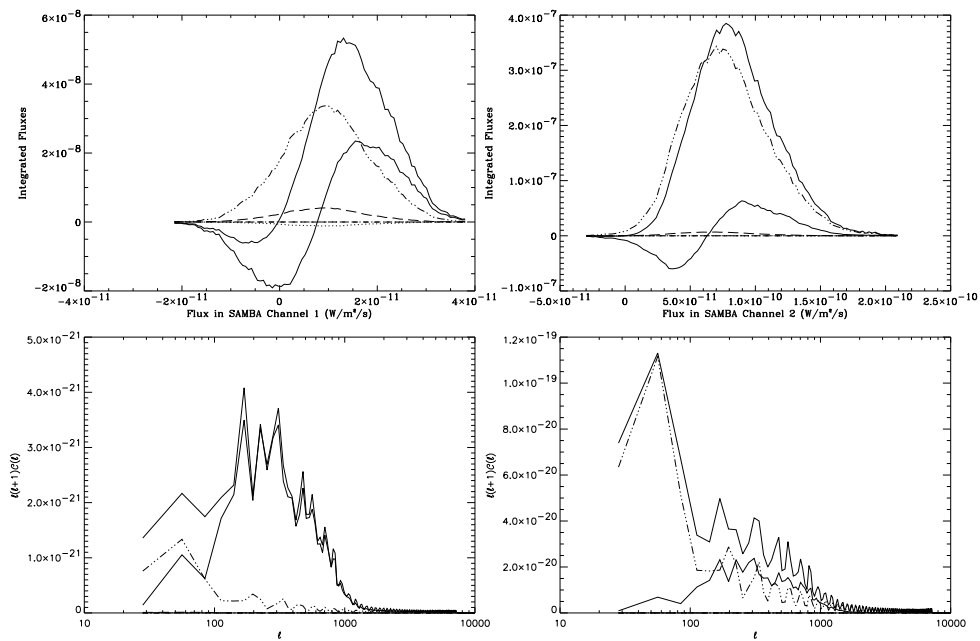

Fig. 2. The top panel shows the flux integrated over all pixels as a function of the total flux in a pixel, for the channels 1 (left) and 2(right). The total, i.e. the x-axis value times the number of pixels with that flux, is given by the thick line. The CMB, $y - SZ$, galaxies, synchrotron, and dust contributions are denoted by a solid, dotted, dashed, dashed–triple–dotted line (the other components being undistinguishable from 0 on this plot). The corresponding angular contributions $\ell(\ell+1)C(\ell)$ are plotted below. The channels 3 & 4 are entirely dust–dominated.

$z$ distribution of galaxies. For lower–$z$ sources of fluctuations, like those coming from the dust in galaxies, we cannot do much more at this time than to extrapolate the dust emission from the IRAS measurements. But there might very well be another cold dust population, spatially uncorrelated with the hot one... In any case, the database structure adopted should permit us to incorporate all the latest developments in relevant theoretical modeling and/or new observations, as they appear.

## Acknowledgements

We thank F.-X. Désert, M. Giard, F. Pajot, M. Pérault, and J.-L. Puget for fruitful discussions and help in the launching of this project.